\documentstyle[12pt]{article}
\catcode`@=11
\newcount\@tempcntc
\def\@citex[#1]#2{\if@filesw\immediate\write\@auxout{\string\citation{#2}}\fi
  \@tempcnta\z@\@tempcntb\m@ne\def\@citea{}\@cite{\@for\@citeb:=#2\do
    {\@ifundefined
       {b@\@citeb}{\@citeo\@tempcntb\m@ne\@citea\def\@citea{,}{\bf ?}\@warning
       {Citation `\@citeb' on page \thepage \space undefined}}%
    {\setbox\z@\hbox{\global\@tempcntc0\csname b@\@citeb\endcsname\relax}%
     \ifnum\@tempcntc=\z@ \@citeo\@tempcntb\m@ne
       \@citea\def\@citea{,}\hbox{\csname b@\@citeb\endcsname}%
     \else
      \advance\@tempcntb\@ne
      \ifnum\@tempcntb=\@tempcntc
      \else\advance\@tempcntb\m@ne\@citeo
      \@tempcnta\@tempcntc\@tempcntb\@tempcntc\fi\fi}}\@citeo}{#1}}
\def\@citeo{\ifnum\@tempcnta>\@tempcntb\else\@citea\def\@citea{,}%
  \ifnum\@tempcnta=\@tempcntb\the\@tempcnta\else
   {\advance\@tempcnta\@ne\ifnum\@tempcnta=\@tempcntb \else \def\@citea{--}\fi
    \advance\@tempcnta\m@ne\the\@tempcnta\@citea\the\@tempcntb}\fi\fi}
\catcode`@=12
\def\be{\begin{equation}}
\def\ee{\end{equation}}
\def\barr{\begin{array}}
\def\earr{\end{array}}
\def\bea{\begin{eqnarray}}
\def\eea{\end{eqnarray}}
\def\bmath{\begin{displaymath}}
\def\emath{\end{displaymath}}
\def\bq{\begin{quote}}
\def\eq{\end{quote}}

\def\as{\alpha_s}

\def\CF{C_{\scriptscriptstyle F}}
\def\NC{N_{\scriptscriptstyle C}}
\def\Li{\mbox{$\mbox{\rm Li}_2$}}
\def\Frac#1#2{\mbox{$\textstyle{#1\over#2}$}}
\def\eps{\varepsilon}
\def\nn{\nonumber\\}

\def\tw{\theta_{\scriptscriptstyle W}}
\def\ct{\cos\theta}

\def\MZ{M_{\scriptscriptstyle Z}}

\def\Pl{P^\ell}

\def\tint{\int{dy\,dz \over \{(1-y)^2-\xi\}^{3/2}}\:\:}
\def\sxi{\mbox{$\sqrt\xi$}}
\def\w{\mbox{$\sqrt{\textstyle\strut 1-\sxi\:\over\textstyle\strut 1+\sxi\:}$}}
\begin{document}
\thispagestyle{empty}
\begin{flushright}
MZ-TH/95-09 \\[-0.2cm]
FTUV/95-13 \\[-0.2cm]
IFIC/95-13 \\[-0.2cm]
July 1995 \\[-0.2cm]
\end{flushright}
\begin{center}

{\bf{\Large Longitudinal Contribution to the Alignment}}\\[.3cm]
{\bf{\Large Polarization of Quarks Produced in}}\\[.3cm]
{\bf{\Large {\boldmath$e^+e^-$}-Annihilation:
            An {\boldmath$O(\as)$} Effect}}\\[1.25cm]

{\large S.~Groote$^*$, J.G.~K\"orner\footnote{Supported in part
by the BMFT, FRG, under contracts 06MZ730 and 06MZ566,\\
and by HUCAM, EU, under contracts CHRX-CT94-0579}} \\[.4cm]
Institut f\"ur Physik, Johannes Gutenberg-Universit\"at \\
Staudingerweg 7, D-55099 Mainz, Germany. \\[1cm]

{\large M.M.~Tung\footnote[7]{Feodor-Lynen Fellow}} \\[.4cm]
Departament de F\'\i sica Te\`orica, Universitat~de Val\`encia \\
and IFIC, Centre Mixte Universitat Val\`encia --- CSIC, \\
C/ Dr.~Moliner, 50, E-46100 Burjassot (Val\`encia), Spain.
\end{center}
\vspace{1cm}
\centerline {\bf ABSTRACT}
\noindent
We calculate the longitudinal contribution to the alignment
polarization $\Pl$ of quarks produced in $e^+e^-$ annihilation. In the
Standard Model, the longitudinal alignment polarization vanishes at the
Born term level and thus receives its first non-zero contribution from
the $O(\as)$ tree graph process. We provide analytical and numerical
results for the longitudinal alignment polarization of massless and
massive quarks, in particular for the recently discovered top quark. \\[1cm]

\section{Introduction}
It is well-known that quarks produced in $e^+e^-$-annihilation are
polarized to a substantial degree. In the absence of beam polarization
effects, the polarization of the quarks results from the interplay of the
parity-violating and parity-conserving pieces of the $Z$-exchange and
$\gamma$-exchange contributions. The degree of polarization of the quarks
depends on the type of the produced quarks and their masses, on the c.m.\
energy (due to the interplay of $\gamma$- and $Z$-exchange contributions),
on beam orientation effects, and on beam polarization effects.

Concerning the detectability of the quark's polarization the case of the
$t$-quark is simplest. The top quark is so heavy that it decays before
hadronizing. Thus, the polarization of the top quark can be directly
studied through an analysis of its subsequent decay modes
\cite{align1,align2}. In the charm and bottom sector, the quark's
polarization is expected to be partially transmitted to the fragmentation
product when the $c$- and $b$-quarks hadronize to the $\Lambda_c$- and
$\Lambda_b$-baryons, respectively, where the polarization transfer is 100\%,
at least in the heavy quark limit \cite{align3}. The issue of the amount
of polarization transfer in the finite mass case is an important and
interesting issue by itself \cite{align4}. Again, the $\Lambda_c$ or
$\Lambda_b$ polarizations can be determined by their subsequent decays
\cite{align5}.

Predictions for the polarization at the Born term level have been worked
out some time ago \cite{align1}. First $O(\as)$ radiative corrections
were presented in Ref.~\cite{align6} where the average alignment
polarization $\langle \Pl\rangle_{\ct}$ was calculated. The averaging was
performed with respect to the cosine of the polar beam orientation
angle~$\theta$.\footnote{
The alignment polarization is defined as the component of the polarization
along the line of flight of the particle. It is also frequently referred to
as the longitudinal polarization. Here we choose the phrase ``alignment
polarization'' in order to set it apart from the ``longitudinal'' beam
orientation component discussed in this paper.} A surprising outcome of the
calculation was that the massless quark limit and the radiative correction
calculation do not interchange. The reason is that there is a finite spin-flip
contribution from collinear gluon emission that survives even as
$m_q\rightarrow 0$~\cite{align6,align7}. The corresponding spin-flip
contribution is not present in massless QCD. To put it succinctly, one finds
$\hbox{\rm QCD}(m_q\rightarrow 0)\neq\hbox{\rm QCD}(m_q=0)$
in the perturbative sector. Such anomalous spin-flip contributions had already
been discussed some time ago, in the context of QED~\cite{align8}.

In this paper, we discuss first results on $O(\as)$ radiative
corrections to orientation-dependent polarization effects. We calculate
the $O(\as)$ corrections to the longitudinal component of the quark's
alignment polarization. The longitudinal alignment polarization is of
particular interest since it vanishes at the Born term level in the
Standard Model (SM) due to the fact that the axial-vector current is a
first class current in the SM. Directly connected to this observation is
the statement that there is no anomalous spin-flip contribution to the
longitudinal alignment polarization for vanishing quark masses.

The plan of the paper is as follows. In Sec.~2, we give general expressions
for the polar angle dependence of the alignment polarization of quarks
produced in $e^+e^-$-annihilation. The polarization is written in terms of
polarized and unpolarized helicity structure functions. In Sec.~3, we list
the Born term contributions to the helicity structure functions. Sec.~3
also contains complete results on the one-loop radiative corrections to the
alignment polarization in the case $m_q=0$. These are particularly simple
for the longitudinal alignment polarization. We briefly touch on the role
of anomalous $O(\as)$ spin-flip contributions to the alignment polarization.
Sec.~4 constitutes the main part of the paper and gives analytical and
numerical results on the massive $O(\as)$ corrections to the
longitudinal alignment polarization. We present results on the single and
double differential distributions of the polarized longitudinal structure
function as well as on its totally integrated form. The Appendix contains
our analytical results on the fully integrated structure functions, where we
also outline how these closed form expressions were arrived at.

\section{Orientation dependence of alignment polarization}
In the absence of beam polarization effects, the orientation-dependent
alignment polarization is given by (see e.g.~\cite{align9})
\be\label{eqn1}
\Pl(\ct)=\frac{d\sigma^\ell/d\ct}{d\sigma/d\ct}
  =\frac{N^\ell(\ct)}{D(\ct)},
\ee
where $d\sigma^\ell$ and $d\sigma$ are the polarized and unpolarized
differential production rates, resp., and where
\bea
N^\ell&=&
  \Frac38(1+\cos^2\theta)g_{14}H_U^{4(-)}
  +\Frac34\sin^2\theta\,g_{14}H_L^{4(-)}
  +\Frac34\ct(g_{41}H_F^{1(-)}+g_{42}H_F^{2(-)}),\label{eqn2}\\
D&=&
  \Frac38(1+\cos^2\theta)(g_{11}H_U^{1(+)}+g_{12}H_U^{2(+)})
  \label{eqn3}\nn&&
  +\Frac34\sin^2\theta(g_{11}H_L^{1(+)}+g_{12}H_L^{2(+)})
  +\Frac34\ct\,g_{44}H_F^{4(+)}
\eea
The polar angle $\theta$ defines the polar orientation of the electron beam
relative to the quark's line of flight. We have defined the sum and the
difference of the hadronic tensor with $\lambda_q=\pm\Frac12$ helicities
of the produced quark as
\be
H_{\mu\nu}^{(\pm)}=H_{\mu\nu}(\lambda_q=+\Frac12)\pm
  H_{\mu\nu}(\lambda_q=-\Frac12),
\ee
where the covariant hadronic tensor is defined as usual by
\be\label{eqn4}
H_{\mu\nu}=\sum_{\bar{q}(g)\ \mbox{\scriptsize spins}}
  \langle q\,\bar{q}(g)|j_\mu|0\rangle\
  \langle 0|j_\nu^\dagger|q\,\bar{q}(g)\rangle.
\ee
The spin sum in Eq.~(\ref{eqn4}) does not include the spin of the quark
which one wants to observe. Thus, in the case of two-body production
$e^+e^-\rightarrow q\bar{q}$, the hadronic tensor is a function of
$q^2$ and the spin vector $s$ of the quark. For the three-body final state
in $e^+e^-\rightarrow q\bar{q}g$, the hadron tensor depends in addition
on two Dalitz-plot type variables which will be specified later on. In the
three-body case, the hadron tensor has more structure than indicated in
Eqs.~(\ref{eqn2}) and~(\ref{eqn3}) which, however, disappears after
azimuthal averaging as is always implied in this paper.

In order to exhibit the orientation-dependence of the produced quark, we
have projected onto the helicity structure functions $H_U=H_{++}+H_{--}$,
$H_L=H_{00}$ and $H_F=H_{++}-H_{--}$, where
\be
H_{\lambda_{Z,\gamma}\lambda_{Z,\gamma}}=\varepsilon^\mu(\lambda_{Z,\gamma})\,
H_{\mu\nu}\;{\varepsilon^\nu}^*(\lambda_{Z,\gamma}),
\ee
and $\lambda_{Z,\gamma}$ is the helicity of the current or the gauge boson.
The hadron tensor $H_{\mu\nu}^i$ carries an extra index $i$ (which is
sometimes suppressed) to specify its vector/axial-vector decomposition
according to
\bea\label{decomp}
H_{\mu\nu}^1&=&\Frac12(H_{\mu\nu}^{VV}+H_{\mu\nu}^{AA}),\nn
H_{\mu\nu}^2&=&\Frac12(H_{\mu\nu}^{VV}-H_{\mu\nu}^{AA}),\\
H_{\mu\nu}^4&=&\Frac12(H_{\mu\nu}^{V\!A}+H_{\mu\nu}^{AV}).\nonumber
\eea
Note that one has $H_{\mu\nu}^{V\!A}=H_{\mu\nu}^{AV}$ for the unpolarized
and alignment polarization cases that are relevant for our discussion.

The electroweak coupling factors, finally, are contained in the
coefficients $g_{ij}(s)$. They are given by
\bea
g_{11}&=&Q_f^2-2Q_fv_ev_f Re\chi_Z+(v_e^2+a_e^2)(v_f^2+a_f^2)|\chi_Z|^2,\nn
g_{12}&=&Q_f^2-2Q_fv_ev_f Re\chi_Z+(v_e^2+a_e^2)(v_f^2-a_f^2)|\chi_Z|^2,\nn
g_{14}&=&2Q_fv_ea_f Re\chi_Z-(v_e^2+a_e^2)2v_fa_f|\chi_Z|^2,\\
g_{41}&=&2Q_fa_ev_f Re\chi_Z-2v_ea_e(v_f^2+a_f^2)|\chi_Z|^2,\nn
g_{42}&=&2Q_fa_ev_f Re\chi_Z-2v_ea_e(v_f^2-a_f^2)|\chi_Z|^2,\nn
g_{44}&=&-2Q_fa_ea_f Re\chi_Z+4v_ea_ev_fa_f|\chi_Z|^2,\nonumber
\eea
with the usual notation~\cite{align9}
\bea
\chi_Z(s)&=&\frac{gM_Z^2s}{s-M_Z^2+iM_Z\Gamma_Z},\quad
g\ =\ \frac{G_F}{8\sqrt 2\pi\alpha}\ \approx\ 4.49\cdot 10^{-5}
\mbox{\rm GeV}^{-2}, \\
v_e&=&-1+4\sin^2\tw,\quad a_e\ =\ -1\quad
  \hbox{\rm for leptons},\nn
v_f&=&1-\Frac83\sin^2\tw,\quad a_f\ =\ 1\quad
  \hbox{\rm for up-type quarks ($Q_f=\Frac23$), and}\qquad\\
v_f&=&-1+\Frac43\sin^2\tw,\quad a_f\ =\ -1\quad
  \hbox{\rm for down-type quarks ($Q_f=-\Frac13$)}.\nonumber
\eea

The average alignment polarization of the quark can be obtained from
Eq.~(\ref{eqn1}) by averaging over $\ct$. One has
\be\label{eqn5}
\langle \Pl(\ct)\rangle  =
\frac{g_{14}H_{U+L}^{4(-)}}{
g_{11}H_{U+L}^{1(+)}+g_{12}H_{U+L}^{2(+)}}=:\Pl_{U+L},
\ee
where we have written $H_{U+L}=H_U+H_L$. In order to project out a given
helicity structure function from Eq.~(\ref{eqn1}), one needs to take moment
averages of $\Pl(\ct)$. In our later discussion, we shall be mainly interested
in the longitudinal alignment structure function $H_L^{4(-)}$, which can be
obtained with respect to the moment factor
$(2-5\cos^2\theta)$. One has
\be\label{p-beam}
\langle(2-5\cos^2\theta)\Pl(\ct)\rangle =
\frac{g_{14}H_L^{4(-)}}{
g_{11}H_{U+L}^{1(+)}+g_{12}H_{U+L}^{2(+)}}=:\Pl_L.
\ee

It is important to realize that both the total alignment polarization
$\Pl_{U+L}$ and the longitudinal alignment polarization $\Pl_L$ cover the
range $-1\le\Pl_L\le 1$, since $\Pl_{U+L}$ can be written as a ratio
$(a-b)/(a+b)$ and $\Pl_L$ as a ratio $(a-b)/(a+b+c)$, where $a$, $b$ and~$c$
are transitions probabilities into definite helicity states and are thus
positive definite.

We want to mention that there is a certain amount of simplification of
Eq.~(\ref{eqn1}) in the mass-zero limit, where $H^{VV}=H^{AA}$. In this
case, $\Pl(\ct)$ is given by
\be
\Pl(\ct)=\frac{\strut g_{14}\left\{\,\frac38(1+\cos^2\theta)H_U^{4(-)}
  +\frac34\sin^2\theta H_L^{4(-)}\,\right\}+\frac34\ct g_{41}H_F^{1(-)}}
  {\strut g_{11}\left\{\,\frac38(1+\cos^2\theta)H_U^{1(+)} + \frac34
  \sin^2\theta H_L^{1(+)}\,\right\}+\frac34\ct g_{44}H_F^{4(+)}}\,.
\ee

\section{Born term contribution\hfil\break
  and the {\boldmath$O(\as)$} mass-zero case}
The Born term contributions to the various hadron tensor components can be
easily worked out. If $v=\sqrt{1-4m_q^2/q^2}$ denotes the velocity of the
quark in the c.m.\ system, one finds
\bea
H_U^{4(-)}&=&4\NC\,q^2 v,\\
H_L^{4(-)}&=&0,\label{eqn6}\\
H_F^{1(-)}&=&2\NC\,q^2(1+v^2),\qquad H_F^{2(-)}\ =\ 2\NC\,q^2(1-v^2),\\
H_U^{1(+)}&=&2\NC\,q^2(1+v^2),\qquad H_U^{2(+)}\ =\ 2\NC\,q^2(1-v^2)\\
H_L^{1(+)}&=&\NC\,q^2(1-v^2)\ =\ H_L^{2(+)},\\
H_F^{4(+)}&=&4\NC\,q^2 v,
\eea
where $\NC=3$ takes into account that each quark pair occurs in three
color states. For the sake of completeness we write down the appropriate
two-body $e^+e^-\rightarrow q\bar{q}$ production cross section which
reads
\be
\frac{d\sigma}{d\ct}=\frac{\pi\alpha^2 v}{3q^4}D(\ct),
\ee
where the above $H_{U,L}^{1,2(+)}$ and $H_F^{4(+)}$ have to be substituted
in the denominator function $D(\ct)$ given in Eq.~(\ref{eqn3}).

The vanishing of the longitudinal alignment component $H_L^{4(-)}$ in
Eq.~(\ref{eqn6}) can be understood in many different ways. It is connected
with the vanishing of the axial-vector helicity amplitude
$H_{0;\lambda_q=\pm\frac12,\lambda_{\bar q}=\pm\frac12}^A$, or,
alternatively, with the vanishing of the longitudinal projection of the
axial-vector current matrix element, i.e.
$(p_1-p_2)^\mu\bar u(p_1)\gamma_\mu\gamma_5v(p_2)=0$. In partial wave
language, $H_{0;\pm\frac12,\pm\frac12}^A$ is contributed to by the
$LS$-amplitude $L=1$ and $S=0$ ($L=1$, $S=1$ does not couple, because the
corresponding Clebsch-Gordan coefficient
$(LL_zSS_z|J^{\gamma,Z}J_z^{\gamma,Z})=(1010|10)$ vanishes).
However, as is well known from positronium and quarkonium considerations,
the charge conjugation  quantum number of the $L=1$, $S=0$ state is
$C=(-1)^{L+S}=-1$ and thus does not match up with the charge conjugation
$C=+1$ of the axial-vector current. Put in a different language, the Born
term contribution to the longitudinal alignment polarization is a second
class current effect and is thus zero in the SM. The above argument easily
generalizes to the $n$-loop case, where one similarly concludes that its
contributions to the longitudinal alignment polarization vanishes.
Nonvanishing contributions start coming in from three- or more-body final
states as for example from the one-gluon emission graph that will be
discussed further on.

Returning to Eq.~(\ref{eqn1}), one can then calculate the Born term
contribution to $\Pl\!(\ct)$. On the $Z$-peak the polarization is
($r_e=2v_ea_e/(v_e^2+a_e^2)$)
\be\label{eqn7}
\Pl(\ct)=-2\frac{v_fa_fv(1+\cos^2\theta)
  +r_e(v_f^2+a_f^2v^2)\ct}{(v_f^2+a_f^2v^2)(1+\cos^2\theta)
  +v_f^2(1-v^2)\sin^2\theta+4r_ev_fa_fv\ct}.
\ee

The result Eq.~(\ref{eqn7}) agrees with the value of $\Pl(\ct)$
quoted in Ref.~\cite{align1}. Note again the absence of the longitudinal
contribution proportional to $\sin^2\theta$ in the numerator. The mass-zero
case is easily obtained by setting $v=1$ in Eq.~(\ref{eqn7}). One can then
directly calculate the following SM values for the average of the alignment
polarization on the $Z$-peak. The values are
$\langle\Pl(\ct)\rangle_u\approx-69\%$ and
$\langle\Pl(\ct)\rangle_d\approx-94\%$ for up- and down-type
quarks, respectively. By using Eq.~(\ref{eqn7}) it is easy to check that there
is very little mass and orientation dependence of $\Pl(\ct)$ on the
$Z$-peak.

To conclude this section, we shall also list the value of the alignment
polarization $\Pl(\ct)$ in the mass-zero case including
$O(\as)$ radiative corrections. The longitudinal structure functions
$H_L^{(\pm)}$ can easily be calculated since they receive only finite
contributions from the one-gluon emission graphs. The $(y,z)$-Dalitz
plot distribution of the relevant longitudinal tree graph contribution
is~\cite{align9}
\be\label{eqn8}
H_L^{4(-)}(y,z)=H_L^{1(+)}(y,z)=64\pi\NC\CF\as\frac{1-y-z}{(1-y)^2}
\ee
($y=1-2p_1\cdot q/q^2$, $z=1-2p_2\cdot q/q^2$, $p_1=p_q$,
$p_2=p_{\bar q}$). After $z$-integration over the $(y,z)$-Dalitz plot with
the appropriate 4-dimensional integration measure, one finds a flat
$y$-distribution
\be\label{eqn9}
\frac1{16\pi^2}q^2\int\limits_0^{1-y}dz
  \frac{1-y-z}{(1-y)^2}=\frac1{32\pi^2}q^2.
\ee
The remaining $y$-integration in the interval $[0,1]$ is trivial and the
fully integrated contributions to the longitudinal structure functions are
then given by~\cite{align10}
\be\label{eqn10}
H_L^{4(-)}=H_L^{1(+)}=2\NC\CF{\as\over\pi} q^2,
\ee
where as usual $\CF=4/3$ denotes the Casimir operator in the fundamental
representation of the $SU(3)$ color group.
The $O(\as)$ integration for $m_q=0$ in Eq.~(\ref{eqn9}) is deceptively
simple. The corresponding $O(\as)$ calculation for $m_q\neq 0$ discussed
in Sec.~\ref{sect4} turns out to be considerably more complicated.

To calculate the remaining massless unpolarized ($U$) and forward-backward
($F$) structure functions, we need to add the loop contributions. Only the
sum of the tree- and loop-contributions to $H_U$ and $H_F$ are IR/M finite.
Let us simply list the remaining structure functions including the anomalous
contributions to the polarized structure functions $H_U^{4(-)}$ and
$H_F^{1(-)}$, i.e. the finite $O(\as)$ anomalous spin flip contributions
that remain as $m_q\rightarrow 0$.
\bea
H_U^{4(-)}&=&\NC\CF{\as\over\pi}(1-[2])q^2,\qquad
H_U^{1(+)}\ =\ \NC\CF{\as\over\pi}q^2,\\
H_F^{1(-)}&=&\NC\CF{\as\over\pi}(0-[2])q^2,\qquad
H_F^{4(+)}\ =\ 0.
\eea
The anomalous spin-flip contributions are listed in the square brackets.
The polarized structure function $H_U^{4(-)}$ can be obtained
from~\cite{align6} (where $U+L$ was calculated) and Eq.~(\ref{eqn10}) using
the relation $U=(U+L)-L$. The forward-backward structure function $H_F^{4(+)}$
has been calculated in~\cite{align11} for the massive quark case and
in~\cite{align12} for the mass-zero case. Since the anomalous spin-flip
contributions cancel in the {\em sum\/} of helicity transitions, the limit
$m_q\rightarrow 0$ is smooth for the unpolarized structure functions. On
the other hand, since $H_U^{4(-)}=H_U^{1(+)}$ and $H_F^{1(-)}=H_F^{4(+)}$
in massless QCD, the normal (unbracketed) no-flip contributions to the
polarized structure functions $H_U^{4(-)}$ and $H_F^{1(-)}$ can immediately
be obtained from the respective unpolarized cases. Thus, e.g. the normal
no-flip contribution to $H_F^{1(-)}$ vanishes since $H_F^{4(+)}$ vanishes.
The total contribution to $H_F^{1(-)}$ including the anomalous spin-flip
term has been calculated in~\cite{align13}. Finally, as has been emphasized
before, there are no $O(\as)$ anomalous spin-flip contributions to the
longitudinal polarized structure function $H_L^{4(-)}$, since these are
associated with collinear gluon emission which are absent in $H_L^{4(-)}$
due to the fact that $H_L^{4(-)}$ vanishes at the Born term level.

\section{{\boldmath$O(\as)$} correction to the longitudinal\hfil\break
  alignment polarization for {\boldmath$m_q\neq 0$}\label{sect4}}
The calculation of the massive $O(\as)$ corrections to the longitudinal
alignment polarization is straightforward but tedious. One first projects out
the relevant longitudinal components and then performs the phase-space
integration to obtain the longitudinal polarized structure functions. We
mention that we define the longitudinal structure functions with respect to
the line of flight of the quark.

The relevant normalized covariant longitudinal projector is thus given by
  \be\label{proj}
  \Pi^{\mu\nu}_L = \eps^\mu(0)\,{\eps^*}^\nu(0) = {1\over p_{1z}^2}\,
  \left(\,p_1^\mu-{p_1\!\cdot\!q\over q^2}\,q^\mu\,\right)
  \left(\,p_1^\nu-{p_1\!\cdot\!q\over q^2}\,q^\nu\,\right),
  \ee
where $p_1$ is the quark's momentum and $q$ denotes the total four-momentum
carried by the gauge boson. In terms of the $(y,z)$-variables, the
longitudinal projector takes the form
\be
  \Pi^{\mu\nu}_L={4\over q^2\left[(1-y)^2-\xi\right]}\,
  \left\{\,p_1^\mu-\Frac{1}{2}(1-y)\,q^\mu\,\right\}
  \left\{\,p_1^\nu-\Frac{1}{2}(1-y)\,q^\nu\,\right\}.
\ee
A straightforward calculation of the $O(\as)$ tree graph contribution gives
\bea
  H_L^{4(-)}(y,z)&=&\Pi^{\mu\nu}_L\,H_{\mu\nu}^{4(-)}(tree)\nn
  &=&\frac{8\pi\as\NC\CF}{\left\{(1-y)^2-\xi\right\}^{3/2}}\,
  \Biggl[\,4(2-3\xi+\xi^2)-(16-8\xi-3\xi^2)y\nn&&
  +8(1+\xi)y^2-(8-10\xi-\xi^2)z+4(2+\xi)yz\nn&&
  -2\xi(1-\xi)\left({y\over z}+{z\over y}\right)
  -\xi(2-3\xi){y^2\over z}+4\xi {y^3\over z}
  +\xi^2 {y^3\over z^2}\,\Biggr].\label{eqn11}
\eea
for the longitudinal projection of the vector/axial-vector interference part
of the polarized hadron tensor. We have used the abbreviation
$\xi=4m_q^2/q^2$. In calculating the squared matrix element
Eq.~(\ref{eqn11}), we have made use of the covariant form of the alignment
polarization $4$-vector
$s_\mu^\ell=\xi^{-{1\over 2}}\,\Big(\sqrt{(1-y)^2-\xi},\,0,\,0,\,1-y\,\Big)$.
In the mass-zero limit $\xi\rightarrow 0$, Eq.~(\ref{eqn11}) can be seen to
collapse to the mass-zero case Eq.~(\ref{eqn8}). Since there are no IR
singularities when integrating $H_L^{4(-)}$, one does not have to introduce
a gluon mass as infrared regulator. Note that the denominator factor
$\{(1-y)^2-\xi\}^{3/2}$ in Eq.~(\ref{eqn11}) vanishes at the edge of phase
space where $p_{1z}=0$ ($\widehat{=}\ y=1-\sqrt\xi$, see Fig.~1). This
vanishing, however, does not correspond to a true singularity but is
canceled by the numerator factor. In fact, $H_L^{4(-)}$ vanishes at
$y=1-\sqrt\xi$. We shall have to return to this point when doing the
$y$-integration later on.

The integration measure for the $(y,z)$ integration appropriate for the
integration of the three-body phase-space is given by
\be\label{eqn12}
dPS=\frac{q^2}{16\pi^2v}\;dy\,dz,
\ee
where the phase-space region is shown in the Dalitz plot in Fig.~1. The
boundary functions $z_\pm(y)$ of the phase space integration are listed in
the Appendix together with the boundary values $y_\pm$. Note that the
phase-space factor Eq.~(\ref{eqn12}) is given by the ratio of three-body and
two-body phase-space factors, since we are giving our results in terms of
the two-body structure functions. For definiteness, we list the appropriate
three-body production rate, which is given by
\vspace{-0.3cm}
\be
\frac{d\sigma}{dy\,dz\,d\ct}
  =\frac{\alpha^2}{48\pi q^2}D(y,z,\ct)
\ee
where the denominator function Eq.~(\ref{eqn3}) is now written in terms of
the $(y,z)$-dependent three-body structure functions $H_{U,L,F}^{(+)}(y,z)$.

As a first step, we perform the $z$-integration to obtain the differential
$y$-distribution of the polarized longitudinal structure function
$H_L^{4(-)}(y)$. One finds
\bea
H_L^{4(-)}(y)&=&\frac{q^2}{16\pi^2v}\int_{z_-(y)}^{z_+(y)}dz\,
  H_L^{4(-)}(y,z)\label{eqn13}\\
  &=&\frac{\as\NC\CF\,q^2y}{2\pi v\{(1-y)^2-\xi\}^{3/2}}\Bigg[
  -\xi\Bigg\{2(1-\xi)+(2-3\xi)y-4y^2\Bigg\}
  \ln\left(\frac{z_+(y)}{z_-(y)}\right)\nn&&
  +\left\{\frac{(4-\xi)^2(2-\xi)}{2(4y+\xi)}
  -\Frac12(16-6\xi-\xi^2)+2(2+5\xi)y\right\}\sqrt{(1-y)^2-\xi}\ \Bigg].
  \nonumber
\eea
The vanishing of $H_L^{4(-)}$ at the soft photon point $y=0$ is explicit in
Eq.~(\ref{eqn13}). Also, as mentioned before, $H_L^{4(-)}$ does not become
singular at $y_{\mbox{\scriptsize max}}=1-\sqrt\xi$, but vanishes at this
point. For $\xi=0$, one recovers the massless result Eq.~(\ref{eqn10}).

As a next step, we do the $y$-integration. As just mentioned,
Eq.~(\ref{eqn13}) is well-behaved for $y\rightarrow 1-\sqrt\xi$,
since the denominator zero is canceled by the numerator factor. However,
when integrating Eq.~(\ref{eqn13}) term by term, one does encounter
spurious singularities in intermediate steps of the calculation. These
spurious singularities will be regularized by slightly deforming the
phase-space edge from $y_+=1-\sqrt\xi$ to $y_+=1-\sqrt\xi-\eps$ (see Fig.~1).
The spurious divergences must and do cancel out after all terms are
recombined. A complete list of analytical expressions for the necessary
spin-dependent integrals encountered in the integration of Eq.~(\ref{eqn13})
is given in the Appendix. Finally, the totally polarized longitudinal
structure function is given by
\bea\label{eqn14}
  H^{4(-)}_L&=&\int dPS\,H_L^{4(-)}(y,z)\ =\ \int dy\,H_L^{4(-)}(y)\nn
  &=&\frac{\as\NC\CF q^2}{2\pi v}\Biggl[\,4(1-\xi)(2-\xi)T_1-
  (16-8\xi-3\xi^2)T_2+8(1+\xi)T_3 \nn
  &&\hskip2.7cm -(8-10\xi-\xi^2)T_4+4(2+\xi)T_5-2\xi(1-\xi)(T_6+T_7)\nn
  &&\hskip2.7cm -\xi(2-3\xi)T_8+4\xi T_9+\xi^2T_{10}\,\Biggr],
  \eea
where the spurious $\ln\eps$ singularities can be seen to cancel in
Eq.~(\ref{eqn14}). Taking the $m_q\rightarrow 0$ limit in Eq.~(\ref{eqn14})
requires some care, but one can check that Eq.~(\ref{eqn14}) reduces to the
massless result Eq.~(\ref{eqn10}) as $m_q\rightarrow 0$.

In Figs.~2, the numerical $O(\as)$ results of the longitudinal alignment
polarization and the total cross section are displayed for $e^+e^-\to\gamma$,
$Z\to t\bar{t}$ with a top-quark mass of $m_t(m_t)=174$ GeV. We have
considered the running of the strong coupling~\cite{align14} by evolving
$\as^{(5)}(\MZ)=0.123$ within 5 flavor QCD to the heavy-quark threshold
and then connecting the threshold value to a theory with 6 active flavors.
Evidently, it suffices to use the QCD renormalization group equations
up to one-loop order. For the bottom quark case to be discussed next,
similar matching conditions relate the couplings of the theory with 5
quarks and an effective theory with only 4 quarks.

The top quark production cross section shown in Fig.~2(a) rises steeply from
threshold at 348 GeV and then settles down to its asymptotic $1/q^2$
behaviour. The longitudinal alignment polarization $\Pl_L$ shown in Fig.~2(b)
is negative and remains quite small at less than $1\%$. It rises from
threshold and has only reached one-third of its asymptotic value
$\Pl_L\simeq\frac{2g_{14}}{3g_{11}}\frac{\as}{\pi}(1+\frac{\as}{\pi})^{-1}
\cong-0.64\%$ (modulo logarithmic effects from the running of
the coupling constant) at 1000 GeV. In Fig.~3, we show the $y$-distribution of
the longitudinal alignment polarization for three different $q^2$-values.
The polarization peaks toward maximal $y$-values, i.e. the polarization is
largest for the lower top quark energies where the production cross section
is smallest. Note that the $y$-distribution of the polarization is defined
by the $O(\as)$ tree graph distribution alone, i.e. the polarization
distribution in Fig.~3 is an $O(\as^0)$ effect, since the QCD coupling
factor~$\as$ cancels in the numerator and denominator. This explains the
sizable polarization values in Fig.~3 as compared to the total $O(\as)$
polarization values in Fig.~2(b). Note also that the soft-gluon point $y=0$
does not require special attention because of the afore-mentioned absence
of an IR singularity in the polarized longitudinal structure function. The
total rate factor in the denominator for Eq.~(\ref{eqn1}) does become singular
for $y\rightarrow 0$ leading to the rapid vanishing of the longitudinal
alignment polarization as $y\rightarrow 0$.

The results for bottom-quark production are presented in Figs.~4. On the
$Z$-peak, we find for the longitudinal beam alignment polarization a
maximal value of $P^\ell_L=-2.05\%$. In the range from 300 GeV to 1000 GeV,
$P^\ell_L$ decreases slightly from $-1.4\%$ to $-1.2\%$. This
fall-off is mainly due to the logarithmic fall-off of $\as$. If one neglects
this effect again, the polarization would increase and reach the asymptotic
value $\Pl_L\approx -1.64\%$. We do not present any $y$-distributions of the
longitudinal alignment polarization for the $b$-quark case since we have
not implemented any collinear cuts in the present calculation. In the
absence of such cuts, the $y$-distribution of the polarization is nominally
zero when $m_b\rightarrow 0$ because the denominator in the polarization
expression is singular for any value of $y$ in this case, due to the
collinear singularity at $z=0$.

We mention that, in our numerical evaluations, we have made use of the
simple Schwinger-type approximations derived in Ref.~\cite{align15} for the
unpolarized cross section, which appear in the denominator of the
polarization expression Eq.~(\ref{eqn5}).
\vskip1cm\noindent
{\bf Acknowledgements.} M.M.T.\ wishes to thank J.~Bernab\'eu,
J.~Pe\~narrocha and A.~Pich for stimulating discussions and gratefully
acknowledges the support given by the Alexander-von-Humboldt Foundation
and the Generalitat Valenciana.

\newpage
\section*{Appendix:\hskip4mm{\boldmath$q\bar{q}g$} Phase-Space Integration}
\setcounter{equation}{0}
\def\theequation{A\arabic{equation}}
The boundaries of the three body phase-space are given by the functions
\be
z_\pm = {2y\over 4y+\xi}\left\{\,1-y-\Frac12\xi\pm\sqrt{(1-y)^2-\xi}\,
\right\},
\ee
where the extremal values of $y$ are
\be
y_- = 0,\qquad y_+\ =\ 1-\sqrt\xi.
\ee
The boundaries of the three-body phase-space are shown in Fig.~1. One first
does the $z$-integration in the above limits. The result is listed in
Eq.~(\ref{eqn13}). As mentioned before, one encounters spurious divergences
at the upper boundary $y=y_{\mbox{\scriptsize max}}$ in individual terms when
further integrating Eq.~(\ref{eqn13}) over $y$. These spurious divergencies
are regularized by introducing the cutoff $\eps$. Then the spin-dependent
integrals with no logarithms have the form ($n=0,1$; $m=1,2$)
\be\label{eqn15}
T(n,m)=\int\limits^{1-\sqrt{\xi}-\eps}_0\!\!\!dy\:
  \frac{y^m}{(4y+\xi)^n((1-y)^2-\xi)}\;.
\ee
It is not difficult to see by complete induction that the following
relation holds
\be
4\,T(n,m) = T(n-1,m-1)-\xi\,T(n,m-1).
\ee
With a short list of the basic integral solutions, one can then successively
construct all the remaining integrals.

The second type of integral needed to integrate Eq.~(\ref{eqn13}) involves
the additional factor $\ln[z_+(y)/z_-(y)]/\sqrt{(1-y)^2-\xi}$ in the
integrand of Eq.~(\ref{eqn15}) where now $n=0$ and $m=1,2,3$. Using again
an appropriate substitution, all the solutions containing also
dilogarithms~\cite{align16} can be found in a straightforward manner.
A complete list of integrals encountered in the spin-dependent calculation
is the following:
%
      \bea
     T_1 & = & \tint \nonumber \\
     & = &
     {2(1-\sxi)\over\sxi(2-\sxi)^2}
     \Bigg[\,-\ln\eps+\Frac{1}{2}\ln\xi-\ln(1+\sxi)+\ln(1-\sxi)+\ln2\,\Biggr]
     \nonumber \\ &&
     +{4\xi\over(4-\xi)^2}
     \Biggl[\,\Frac{3}{2}\ln\xi-\ln(1+\sxi)-2\ln(2-\sxi)+\ln2\,\Biggr],
     \label{app1}\\
     \nonumber \\
     \nonumber \\
     T_2 & = & \tint y \nonumber \\
     & = &
     {2(1-\sxi)^2\over\sxi(2-\sxi)^2}
     \Bigg[\,-\ln\eps+\Frac{1}{2}\ln\xi-\ln(1+\sxi)+\ln(1-\sxi)+\ln2\,\Biggr]
     \nonumber \\ &&
     -{\xi^2\over(4-\xi)^2}
     \Biggl[\,\Frac{3}{2}\ln\xi-\ln(1+\sxi)-2\ln(2-\sxi)+\ln2\,\Biggr]
     +\Frac{1}{2}\ln\xi
     \nonumber \\ &&
     -\ln(1+\sxi)+\ln2, \\
     \nonumber \\
     \nonumber \\
     T_3 & = & \tint y^2 \nonumber \\
     & = &
     {2(1-\sxi)^3\over\sxi(2-\sxi)^2}
     \Bigg[\,-\ln\eps+\Frac{1}{2}\ln\xi-\ln(1+\sxi)+\ln(1-\sxi)+\ln2\,\Biggr]
     \nonumber \\ &&
     +{\xi^3\over4(4-\xi)^2}
     \Biggl[\,\Frac{3}{2}\ln\xi-\ln(1+\sxi)-2\ln(2-\sxi)+\ln2\,\Biggr]
     \nonumber \\ &&
     +\Frac{1}{4}(8-\xi)\Biggl[\,\Frac{1}{2}\ln\xi-\ln(1+\sxi)+\ln2\,\Biggr]
     +1-\sxi, \\
     \nonumber \\
     \nonumber \\
     T_4 & = & \tint z \nonumber \\
     & = &
     2\,{(1-\sxi)^2\over(2-\sxi)^3}
     \Bigg[\,-\ln\eps+\Frac{1}{2}\ln\xi-\ln(1+\sxi)+\ln(1-\sxi)+\ln2\,\Biggr]
     \nonumber \\ &&
     +{\xi(32-20\,\xi-\xi^2)\over 2(4-\xi)^3}
     \Biggl[\,\Frac{3}{2}\ln\xi-\ln(1+\sxi)-2\ln(2-\sxi)+\ln2\,\Biggr]
     \nonumber \\ &&
     +2\,{\xi(1-\sxi)\over(4-\xi)(2-\sxi)^2}
     -\Frac{1}{2}\left[\,\Frac{1}{2}\ln\xi-\ln(1+\sxi)+\ln2\,\right], \\
     \nonumber \\
     \nonumber \\
     T_5 & = &
     \tint y\,z \nonumber \\
     & = &
     2\,{\left({1-\sxi\over2-\sxi}\right)^3}\;
     \Bigg[\,-\ln\eps+\Frac{1}{2}\ln\xi-\ln(1+\sxi)+\ln(1-\sxi)+\ln2\,\Biggr]
     \nonumber \\ &&
     -{\xi^2(12-7\,\xi)\over 2(4-\xi)^3}
     \Biggl[\,\Frac{3}{2}\ln\xi-\ln(1+\sxi)-2\ln(2-\sxi)+\ln2\,\Biggr]
     \nonumber \\ &&
     +{2\over 4-\xi}\Biggl[\,\xi\left({1-\sxi\over2-\sxi}\right)^2+
     \sxi-1\,\Biggr]
     -\Frac{1}{2}\left[\,\Frac{1}{2}\ln\xi-\ln(1+\sxi)+\ln2\,\right], \\
     \nonumber \\
     \nonumber \\
     T_6 & = &
     \tint \frac yz \nonumber \\
     & = &
     2\,{1-\sxi\over\xi(2-\sxi)}
     \Bigg[\,-\ln\eps+\Frac{1}{2}\ln\xi-\ln(1+\sxi)+\ln(1-\sxi)
     +\ln2+2\,\Biggr]
     \nonumber \\ &&
     -{4-3\,\xi\over\xi(4-\xi)}
     \Biggl[\,\Frac{3}{2}\ln\xi-\ln(1+\sxi)-2\ln(2-\sxi)+\ln2\,\Biggr]
     \nonumber \\ &&
     -{1\over\xi}
     \left[\,\Frac{1}{2}\ln\xi-\ln(1+\sxi)+\ln2\,\right]
     -{2v\over\xi}\ln\left(\frac{1+v}{1-v}\right), \\
     \nonumber \\
     \nonumber \\
     T_7 & = &
     \tint \frac zy \nonumber \\
     & = &
     2\,{1-\sxi\over(2-\sxi)^3}
     \Bigg[\,-\ln\eps+\Frac{1}{2}\ln\xi-\ln(1+\sxi)+\ln(1-\sxi)+\ln2\,\Biggr]
     \nonumber \\ &&
     -4\,{8-6\,\xi-\xi^2\over(4-\xi)^3}
     \Biggl[\,\Frac{3}{2}\ln\xi-\ln(1+\sxi)-2\ln(2-\sxi)+\ln2\,\Biggr]
     \nonumber \\ &&
     -8\,\frac{1-\sxi}{(4-\xi)(2-\sxi)^2}, \\
     \nonumber \\
     \nonumber \\
     T_8 & = &
     \tint \frac{y^2}z \nonumber \\
     & = &
     2\,{(1-\sxi)^2\over\xi(2-\sxi)}
     \Bigg[\,-\ln\eps+\Frac{1}{2}\ln\xi-\ln(1+\sxi)+\ln(1-\sxi)+\ln2+2\,\Biggr]
     \nonumber \\ &&
     -{4-3\,\xi+\xi^2\over\xi(4-\xi)}
     \Biggl[\,\Frac{3}{2}\ln\xi-\ln(1+\sxi)-2\ln(2-\sxi)+\ln2\,\Biggr]
     \nonumber \\ &&
     +{1+\xi\over\xi}\Biggl[\,-\Frac{1}{2}\ln\xi+\ln(1+\sxi)-\ln2\,\Biggr]
     -{2v\over\xi}\ln\left(\frac{1+v}{1-v}\right)
     \nonumber \\ &&
     +\Frac{1}{2}\ln\left({1+w\over 1-w}\right)^2
     +\Frac{1}{2}\Biggl[\,\Frac{1}{2}\ln\xi-\ln(1+\sxi)+2\ln(2+\sxi)
     -5\ln2\,\Biggr]\ln(1-w^2)
     \nonumber \\ &&
     +\Li\left({1+w\over 2}\right)+\Li\left({1-w\over 2}\right)
     -2\,\Li\left({2\sxi\over 2+\sxi}\right)
     -2\,\Li\left({2+\sxi\over 4}\right)
     \nonumber \\ &&
     -\Frac{1}{6}\pi^2+\ln^2 2
     +\Li\left({2\sxi\over(2+\sxi)(1+w)}\right)
     +\Li\left({2\sxi\over(2+\sxi)(1-w)}\right)
     \nonumber \\ &&
     +\Li\left((2+\sxi){(1+w)\over 4}\right)
     +\Li\left((2+\sxi){(1-w)\over 4}\right), \\
     \nonumber \\
     \nonumber \\
     T_9 & = &
     \tint {y^3\over z} \nonumber \\
     & = &
     2\,{(1-\sxi)^3\over\xi(2-\sxi)}
     \Bigg[\,-\ln\eps+\Frac{1}{2}\ln\xi-\ln(1+\sxi)+\ln(1-\sxi)+\ln2+2\,\Biggr]
     \nonumber \\ &&
     -{16+20\,\xi-12\,\xi^2+\xi^3\over4\,\xi(4-\xi)}
     \Biggl[\,\Frac{3}{2}\ln\xi-\ln(1+\sxi)-2\ln(2-\sxi)+\ln2\,\Biggr]
     \nonumber \\ &&
     +\left(2+\Frac{1}{4}\xi+{1\over\xi}\right)
     \Biggl[\,-\Frac{1}{2}\ln\xi+\ln(1+\sxi)-\ln2\,\Biggr]
     -\left(4+{2\over\xi}\right)v\,\ln\left(\frac{1+v}{1-v}\right)
     \nonumber \\ &&
     +\Frac{3}{2}\ln\left({1+w\over 1-w}\right)^2
     +\Frac{3}{2}\Biggl[\,\Frac{1}{2}\ln\xi-\ln(1+\sxi)+2\ln(2+\sxi)
     -5\ln2\,\Biggr]\ln(1-w^2)
     \nonumber \\ &&
     +3\,\Bigg[\,
     \Li\left({1+w\over 2}\right)+\Li\left({1-w\over 2}\right)
     -2\,\Li\left({2\sxi\over 2+\sxi}\right)
     -2\,\Li\left({2+\sxi\over 4}\right)
     \nonumber \\ &&
     -\Frac{1}{6}\pi^2+\ln^2 2
     +\Li\left({2\sxi\over(2+\sxi)(1+w)}\right)
     +\Li\left({2\sxi\over(2+\sxi)(1-w)}\right)
     \nonumber \\ &&
     +\Li\left((2+\sxi){(1+w)\over 4}\right)
     +\Li\left((2+\sxi){(1-w)\over 4}\right)
     \,\Bigg]+1-\sxi, \\
     \nonumber \\
     \nonumber \\
     T_{10} & = & \tint \frac{y^3}{z^2} \nonumber \\
     & = &
     2\,{(1-\sxi)^2\over\xi^{3/2}}
     \Bigg[\,-\ln\eps+\Frac{1}{2}\ln\xi-\ln(1+\sxi)+\ln(1-\sxi)+\ln2\,\Biggr]
     \nonumber \\ &&
     +{8\over\xi}\,
     \Biggl[\,\Frac{1}{2}\ln\xi-\ln(1+\sxi)+\ln2\,\Biggr]
     +{4\over\xi}\,(1-\sxi).\label{app2} \\ \nonumber
     \eea
We have used the short hand notation \vskip.5cm $$w=\w\qquad.$$
\newpage
\vspace{3cm}

\centerline{\bf\Large Figure Captions }
\vspace{1cm}
\newcounter{fig}
\begin{list}{\bf\rm Fig.\ \arabic{fig}: }{\usecounter{fig}
\labelwidth1.6cm \leftmargin2.5cm \labelsep0.4cm \itemsep0ex plus0.2ex }

\item Graphical representation of the $q\bar{q}g$ phase-space
for massive quarks ($\xi=4m_q^2/q^2\ne 0$) in terms of
the kinematic parameters $y=1-2p_1\!\cdot\!q/q^2$ and $z=1-2p_2\!\cdot q/q^2$.
All boundary curves (shown are $\xi=0.1,0.2,\ldots,0.9$) intersect with the
origin and have their maximum values on the axes at $y_+,z_+=1-\sqrt{\xi}$.

\item The c.m.\ energy dependence of (a) the longitudinal beam alignment
polarization and (b) the total cross section for top-quark production at the
one-loop QCD level. We have taken $m_t(m_t)=174\,\mbox{\rm GeV}$ and
$\as^{(5)}(\MZ)=0.123$.

\item The longitudinal beam alignment polarization $P_L^\ell$ for
$e^+ e^-\to\gamma,Z\to t\bar{t}g$ as a function of the phase-space
parameter $y=1-2p_1\!\cdot\!q/q^2$. The curves correspond
to fixed c.m.\ energies of 350 GeV, 500 GeV and 1000 GeV.

\item (a) Order $\as$ longitudinal beam alignment polarization and (b)
total production rate for the bottom quark $m_b(m_b)=4.3\,\mbox{\rm GeV}$
as a function of the c.m.\ energy $\sqrt{q^2}$.

\end{list}
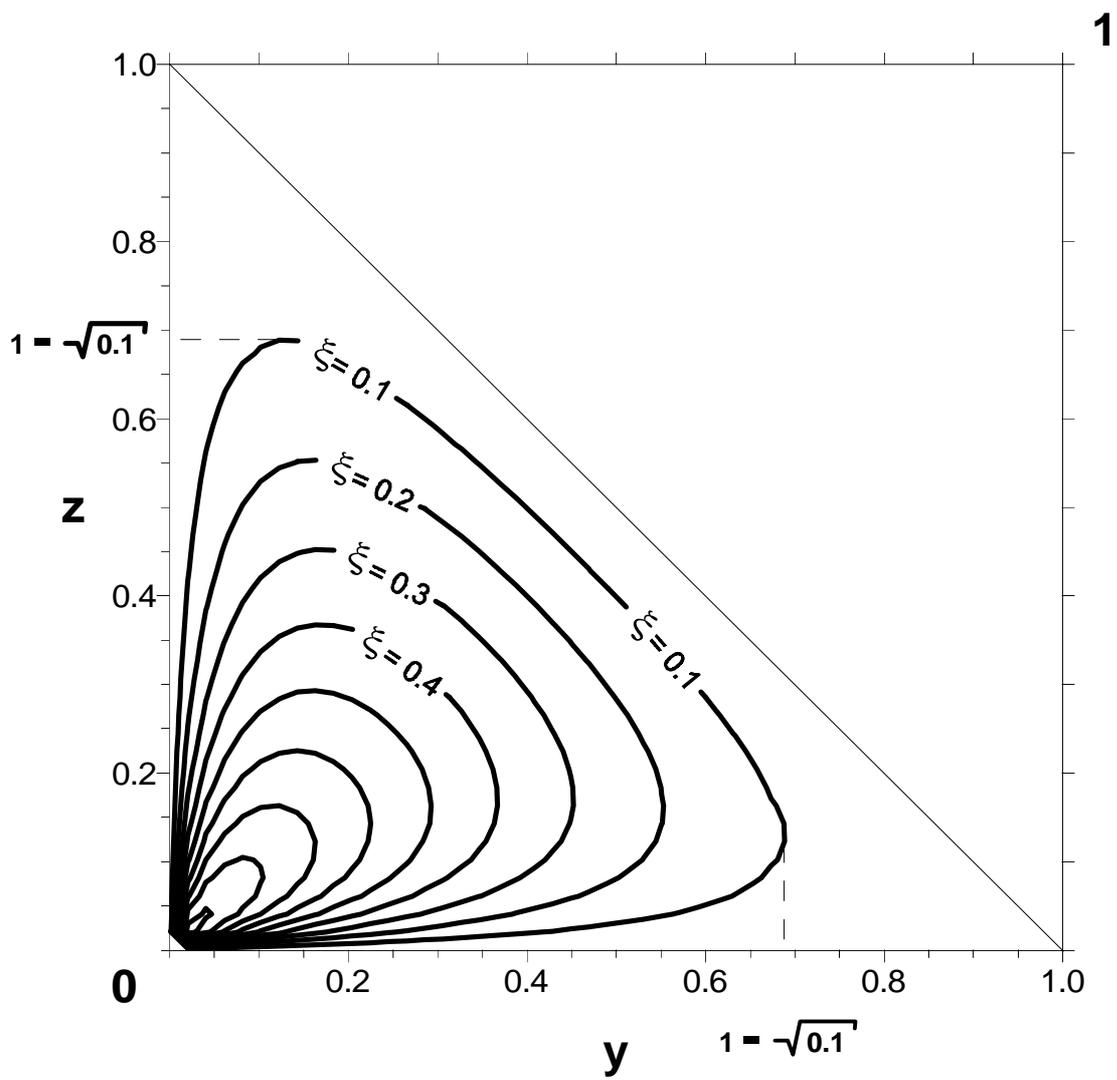
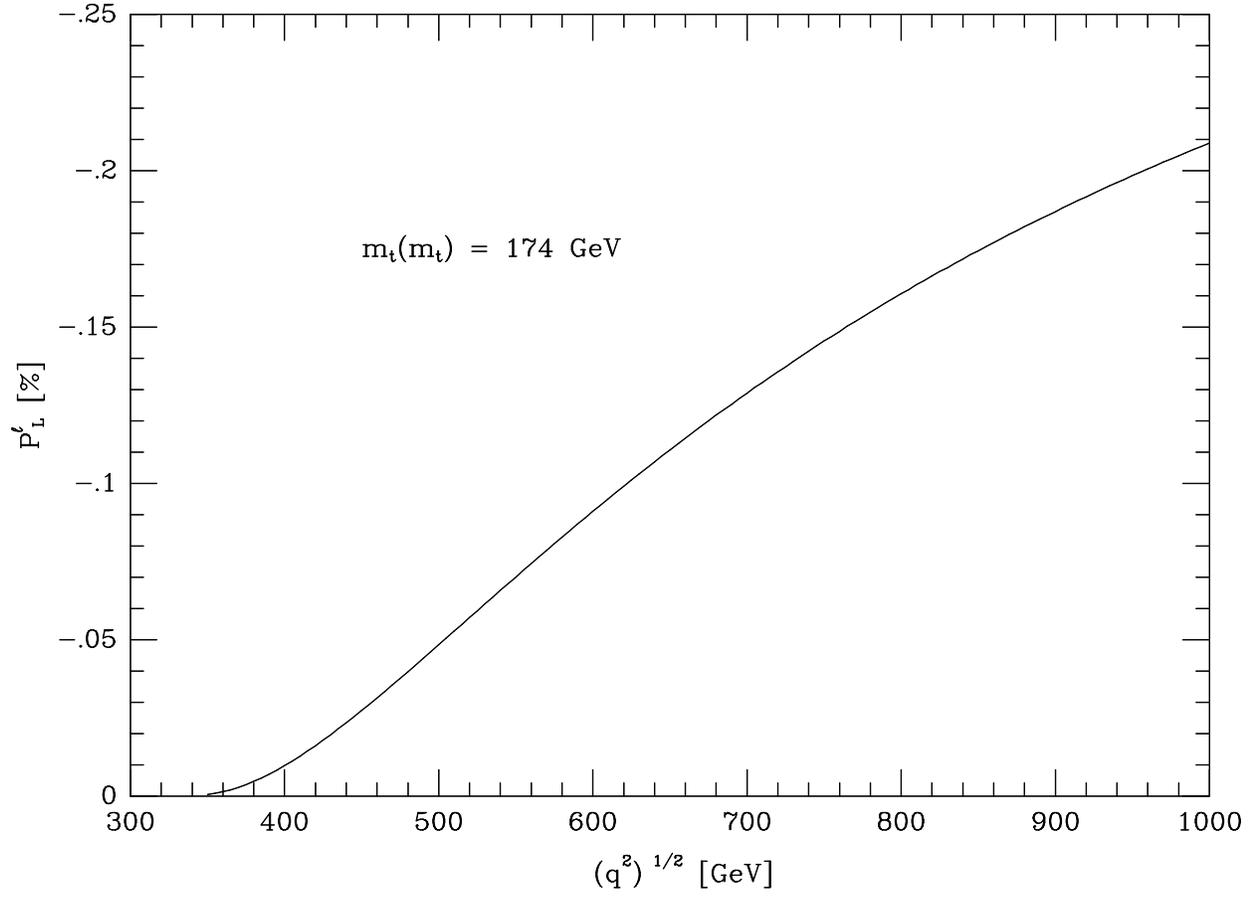
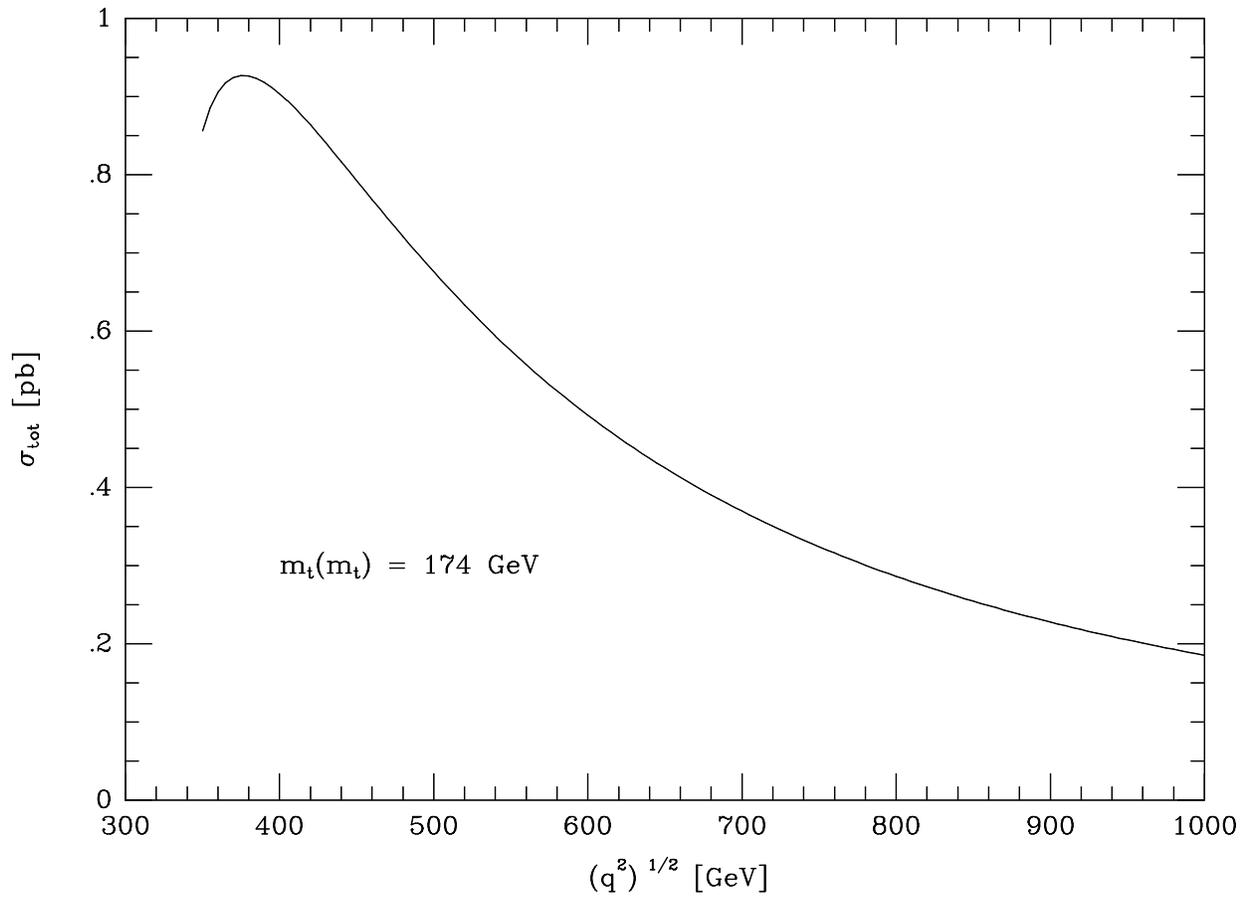
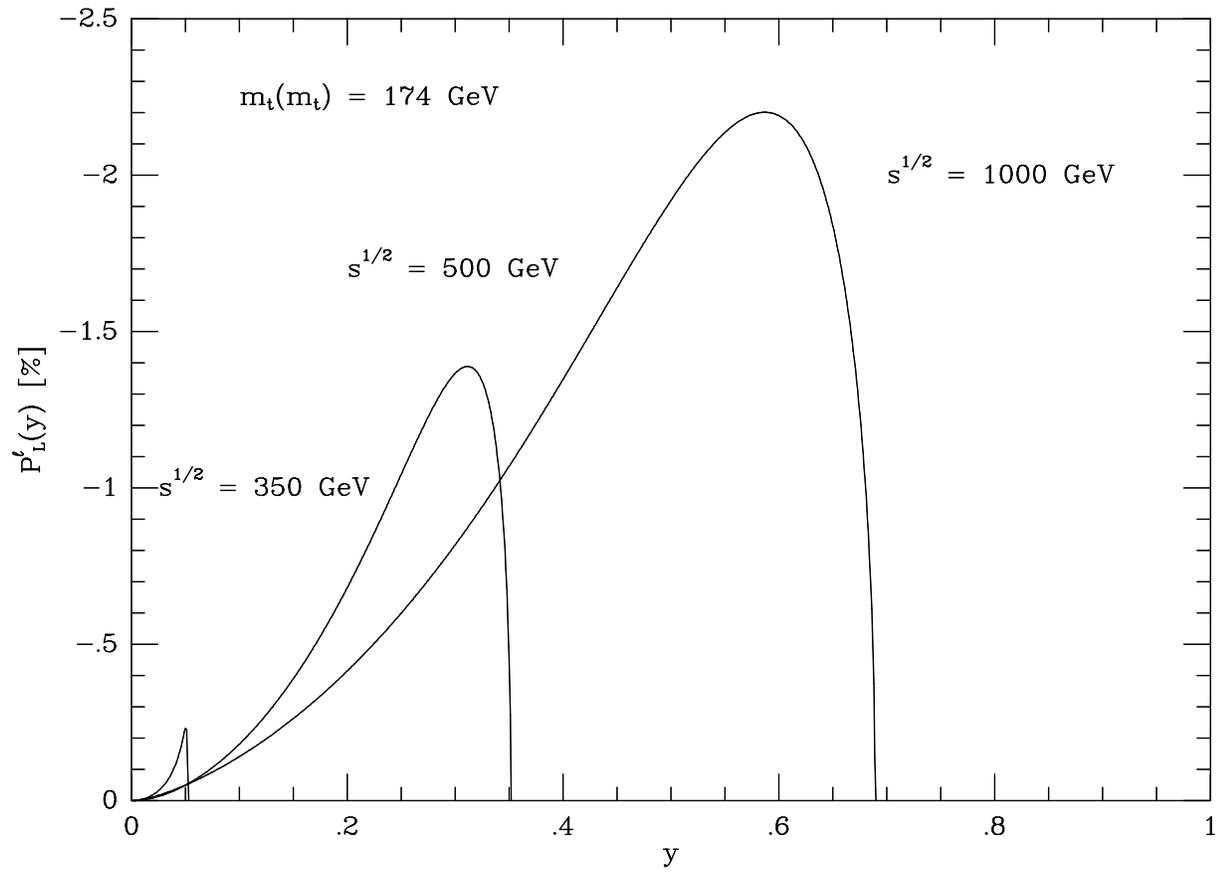
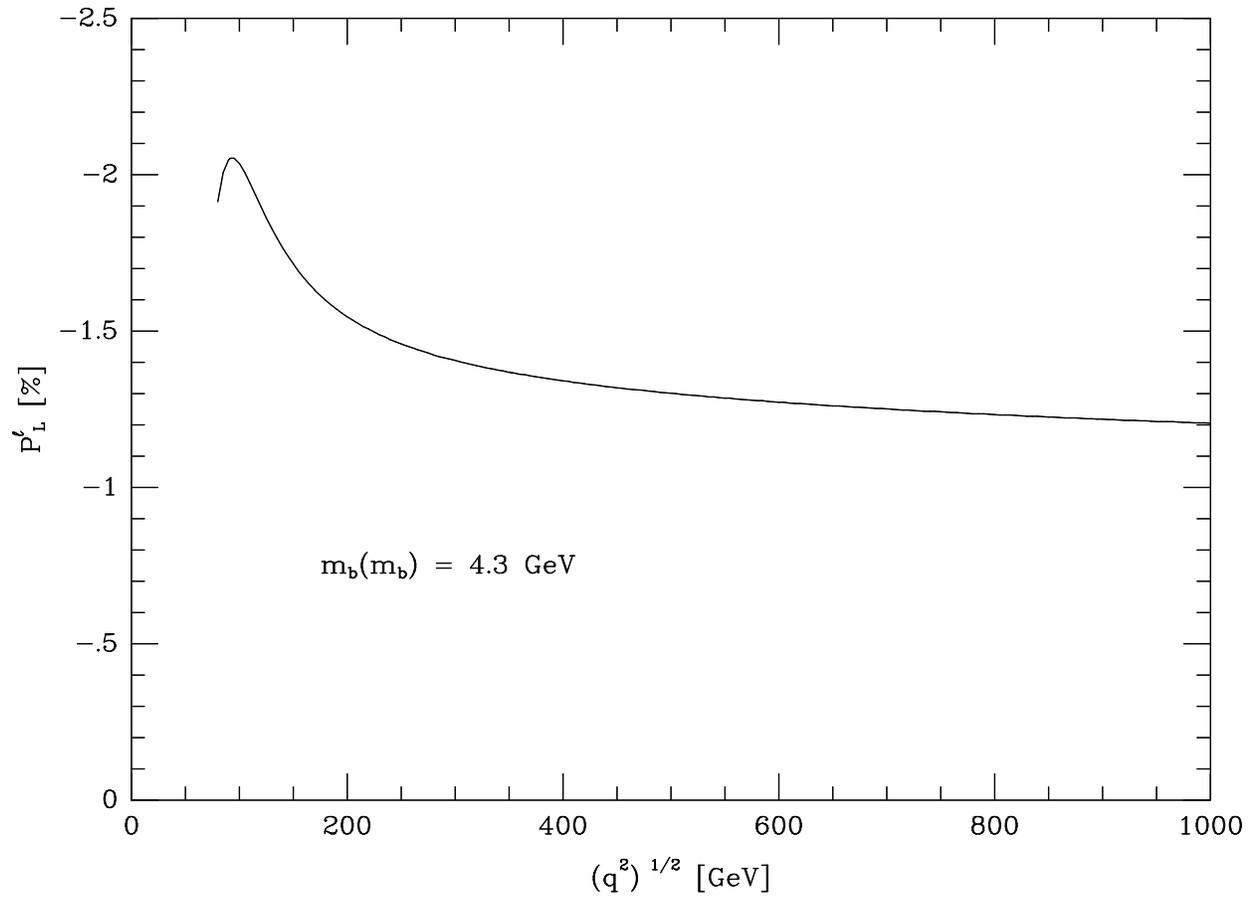
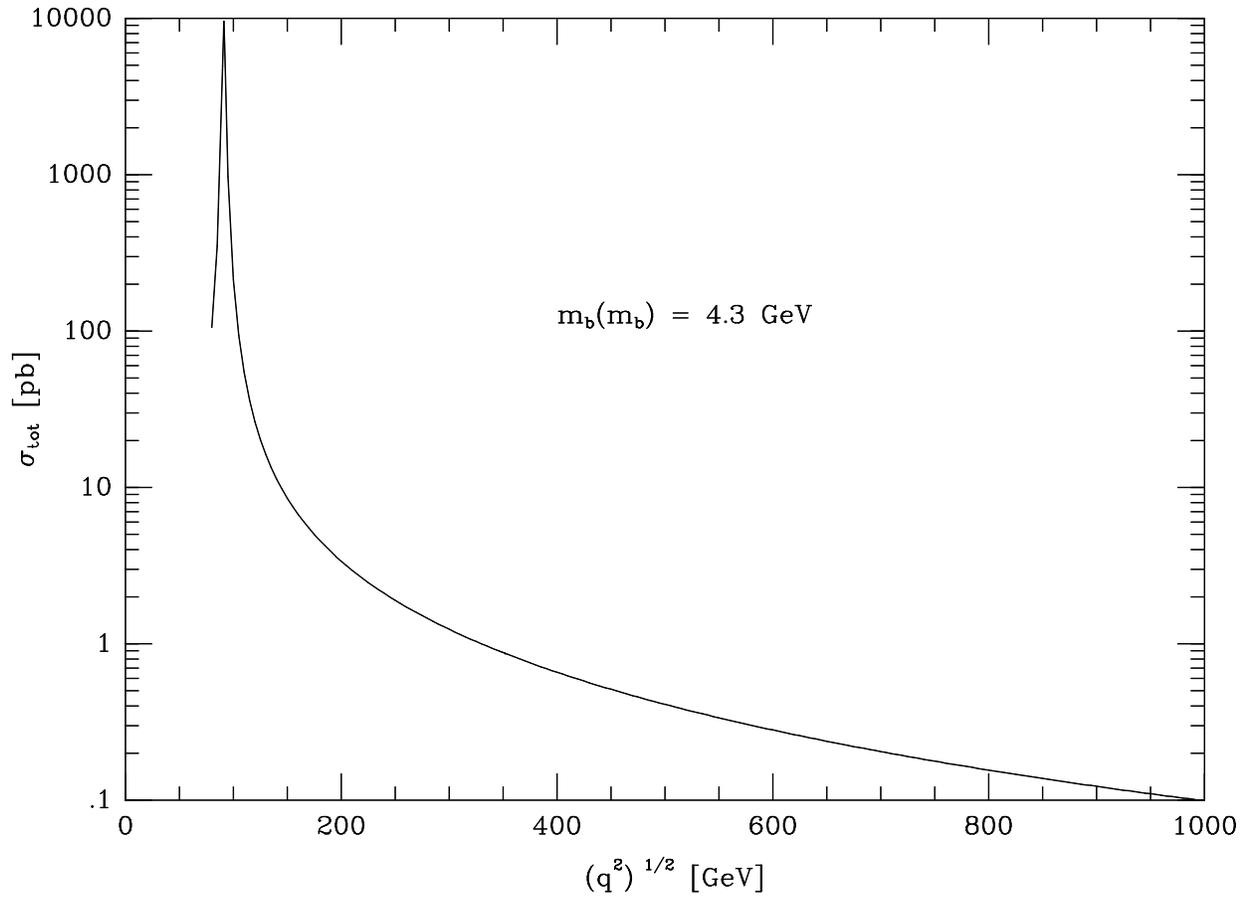
\end{document}